\documentclass[aps,prd,onecolumn,groupedaddress,showpacs,nofootinbib,amssymb]{revtex4}
\usepackage[dvips]{graphicx}
\usepackage{amssymb}
\usepackage{amsmath}
\usepackage{graphicx,color}
\usepackage{amsfonts}
\usepackage{bm}
\usepackage{cancel}
\usepackage{comment}

\newcommand{\be}{\begin{equation}}
\newcommand{\ee}{\end{equation}}
\newcommand{\bea}{\begin{eqnarray}}
\newcommand{\eea}{\end{eqnarray}}
\newcommand{\beaa}{\begin{eqnarray*}}
\newcommand{\eeaa}{\end{eqnarray*}}

\newcommand{\nn}{\nonumber \\}
\newcommand{\e}{\mathrm{e}}

\allowdisplaybreaks[4]
\begin{document}
\title{Unifying Inflation with Early and Late-time Dark Energy in $F(R)$ Gravity}
\author{Shin'ichi Nojiri$^{1,2}$, Sergei D. Odintsov$^{3,4,5,6}$, V.K. Oikonomou,$^{7,5,6}$\,\thanks{v.k.oikonomou1979@gmail.com}}
\affiliation{ $^{1)}$ Department of Physics,
Nagoya University, Nagoya 464-8602, Japan \\
$^{2)}$ Kobayashi-Maskawa Institute for the Origin of Particles
and the Universe,
Nagoya University, Nagoya 464-8602, Japan \\
$^{3)}$ ICREA, Passeig Luis Companys, 23, 08010 Barcelona, Spain \\
$^{4)}$ Institute of Space Sciences (IEEC-CSIC) C. Can Magrans
s/n, 08193 Barcelona, Spain \\
$^{5)}$ International Laboratory for Theoretical Cosmology, Tomsk
State University of Control Systems
and Radioelectronics (TUSUR), 634050 Tomsk, Russia \\
$^{6)}$ Tomsk State Pedagogical University, 634061 Tomsk, Russia\\
$^{7)}$ Department of Physics, Aristotle University of
Thessaloniki, Thessaloniki 54124, Greece}

\begin{abstract}
In this work we shall present models of $F(R)$ gravity which
realize in a unified way the inflationary era along with a
post-inflationary early dark energy era, with the late-time dark
energy era. We shall use two approach methods in order to realize
the unified cosmological eras, firstly we specify a Hubble rate
which may describe the three distinct acceleration eras, and then
by using well known $F(R)$ gravity reconstruction techniques, we
shall find the differential equation which may yield the $F(R)$
gravity that realizes the cosmologies. In our second approach, we
shall present in a qualitative way, several $F(R)$ gravities which
unify the inflationary era with the early and late-time dark
energy eras, and we discuss several qualitative issues related to
the terms that realize the post-inflationary early dark energy
era. We quantify our analysis by numerically solving the Friedmann
equation, using the redshift as the main variable, and expressing
all the physical quantities as functions of the statefinder
$y_\mathrm{H}(z)$, which depends on the redshift and the Hubble
rate $H(z)$. For the model studied numerically, we present the
behavior of some statefinder quantities, like the deceleration
parameter, and we calculate the dark energy density parameter and
the dark energy equation of state parameter at present time. After
demonstrating that the dark energy era is viable, we investigate
when the early dark energy term does not affect the late-time era,
and this restricts the free parameters of the model.
\end{abstract}
\maketitle

\section{Introduction}

The $\Lambda$ Cold Dark Matter ($\Lambda$CDM) model serves as the
most successful model of cosmology up to date, since the Cosmic
Microwave Background (CMB) polarization data coincide to a great
extent with the $\Lambda$CDM predictions \cite{Liddle:2000cg}. The
$\Lambda$CDM model heavily relies on two main components that
control Universe's dynamics, the cosmological constant $\Lambda$
and the assumption of a cold dark matter particle. There are still
many issues to be resolved in the cosmology of $\Lambda$CDM model,
strongly related with the cosmological constant itself and the
dark matter particle. With regard to the cosmological constant,
there is no evidence that it is indeed constant, that is, whether
the Universe is accelerating due to the presence of a cosmological
constant, or a dynamical dark energy evolution is what is actually
observed. With regard to the dark matter component, we really do
not know whether a particle actually constitutes the dark matter
component of the Universe. For the moment, both these issues are
still scrutinized by both theoretical cosmologists and
observational cosmologists. Apart from these issues, there are
also other observational data that challenge theoretical
cosmology, one of which is the so-called $H_0$-tension
\cite{Aylor:2018drw,Wong:2019kwg,Verde:2019ivm,Knox:2019rjx},
which is a mystery. The $H_0$-tension problem casts doubt on the
$\Lambda$CDM model, and this tension is actually the discrepancy
between the observed values of the Hubble rate based on CMB data
\cite{Aghanim:2018eyx}, and the value of the Hubble rate measured
by using low-redshift methods, like the Cepheid variables
\cite{Riess:2016jrr}. The CMB value of the Hubble rate is $H_0=
\left(67.8 \pm 0.9\right)\, \mathrm{km/s/Mpc}$ while the Cepheid
based value is $H_0 = \left( 73.24 \pm 1.74\right)\,
\mathrm{km/s/Mpc}$, so the difference is not small. Apart from the
Cepheid variables measurements, there exist other measurements
\cite{Migkas:2017vir,Ramos-Ceja:2019zxt}, that also verify the
$H_0$-tension. A theoretical proposal that may explain the
$H_0$-tension problem, is the introduction of an early dark energy
era
\cite{Doran:2006kp,Bhattacharyya:2019lvg,Sakstein:2019fmf,Tian:2019enx},
which relies on the prediction of an post-inflationary
acceleration era occurring after the recombination era, so at a
redshift $z\sim 1100$.

Modified gravity
\cite{Nojiri:2017ncd,Nojiri:2009kx,Capozziello:2011et,Capozziello:2010zz,Nojiri:2006ri,
Nojiri:2010wj,delaCruzDombriz:2012xy,Olmo:2011uz}, can
successfully mimic the $\Lambda$CDM model at late-times
\cite{Capozziello:2002rd,Carroll:2003wy,Nojiri:2003ft,Nojiri:2007as,Nojiri:2007cq,Cognola:2007zu,Nojiri:2006gh}
and also at the same time provide a successful mechanism for
generating inflation. It has been shown sometimes ago that
early-time inflation maybe successfully unified with late dark
energy epoch \cite{Nojiri:2003ft}. After that a number of
realistic modified gravities were constructed, in which inflation
is unified with dark energy within the same model, see Refs.
\cite{Nojiri:2007as,Nojiri:2007cq,Cognola:2007zu,Nojiri:2006gh,Appleby:2007vb,Elizalde:2010ts,Odintsov:2020nwm}.
In some $F(R)$ gravity models
\cite{Odintsov:2019evb,Odintsov:2019mlf}, dark matter is composed
by some weakly interacting massive particle
\cite{Bertone:2004pz,Bergstrom:2000pn,Mambrini:2015sia,Profumo:2013yn,Hooper:2007qk,Oikonomou:2006mh}
(WIMP), however it is possible that modified gravity can also
describe dark matter effects in the Universe
\cite{Capozziello:2002rd,Capozziello:2003tk}. Among all modified
gravities, $F(R)$ gravity models are considered to be the most
important ones, since the modification uses functions of the
scalar curvature, the most fundamental geometric-related quantity,
and for a recent review on $F(R)$ gravity cosmological and
astrophysical phenomenology, see for example
\cite{Nojiri:2017ncd}.

In view of the importance of $F(R)$ gravity among modified
gravities, in this work we shall investigate how a
post-inflationary early dark energy era can be realized in the
context of $F(R)$ gravity. Our aim is to unify the inflationary
era with the early and late dark energy eras. We shall investigate
in a quantitative way how an early dark energy era can be
realized, however we shall also provide a qualitative approach, by
constructing an appropriate $F(R)$ gravity model which may unify
all the acceleration eras. Moreover, we shall also study in detail
the phenomenology of one of these models, focusing on the
late-time behavior, and we shall examine the effects of the early
dark energy generating term on the late-time behavior. This will
give us insights on the values of the free parameters of the
model. We shall use the statefinder quantity $y_\mathrm{H}(z)$
which is a function of the redshift and of the Hubble rate $H(z)$,
and by examining the era which corresponds to redshifts
$z=[0,10]$, we shall numerically solve the Friedmann equation in
order to check at first hand the late-time phenomenological
predictions of the model. In addition, we shall also examine the
growth factor and the behavior of the effective gravitational
constant as functions of the redshift, and investigate the effects
of the early dark energy term on these quantities. The results of
the numerical analysis will show us how the free parameters of the
early dark energy term can be chosen in such a way so that it does
not affect the late-time era, but solely the recombination epoch.

We need to stress the conceptual differences of our approach when
compared to the existing literature results
\cite{Doran:2006kp,Bhattacharyya:2019lvg,Sakstein:2019fmf,Tian:2019enx}.
In Refs. \cite{Doran:2006kp,Bhattacharyya:2019lvg,Tian:2019enx}
quintessential scalar fields are used, while in Ref.
\cite{Sakstein:2019fmf} the early dark energy era is generated by
massive neutrinos. In our case, the early dark energy era can be
realized by the presence of appropriately chosen scalar curvature
terms, which during inflation and the late-time acceleration eras
are extremely subdominant, and are chosen to only dominate the
evolution around after the recombination era, so at a redshift
$z\sim 1100$.

\section{Realization of Early Dark Energy with $F(R)$ Gravity: A Quantitative Approach}

Let us first consider how the early dark energy era can be
realized by $F(R)$ gravity by using well known reconstruction
techniques for $F(R)$ gravity. We shall use the notation and
formalism of Ref.~\cite{Nojiri:2009kx}. In order to render the
article self-contained, let us briefly review the reconstruction
procedure for $F(R)$ gravity and how an arbitrary cosmological
evolution can be realized, for more details we refer the reader in
Ref.~\cite{Nojiri:2009kx}. The gravitational action of the $F(R)$
gravity in the presence of perfect matter fluids is,
\begin{equation}
\label{Hm0} S= \int d^4x \sqrt{-g} \left(\frac{F(R)}{2\kappa^2} +
\mathcal{L}_\mathrm{matter} \right) \, ,
\end{equation}
where $\mathcal{L}_\mathrm{matter}$ is the Lagrangian density of
the perfect matter fluids that are present. Assuming the
Friedman-Robertson-Walker (FRW) space-time with flat spatial part,
\begin{equation}
\label{FRWmetric}
ds^2 = - dt^2 + a(t)^2 \sum_{i=1,2,3} \left( dx^i \right)^2 \, ,
\end{equation}
upon varying the action of Eq.~(\ref{Hm0}) with respect to the
metric tensor, we obtain,
\begin{equation}
\label{Hm1}
0 = - \frac{F(R)}{2} + 3 \left( H^2 + \dot H\right) F'(R)
 - 18 (\left(4 H^2 \dot H + H\ddot H\right) F''(R) + \kappa^2 \rho
\, ,
\end{equation}
where the scalar curvature for the FRW Universe is given by
$R=6\dot H + 12 H^2$. We may rewrite Eq.~(\ref{Hm1}) by using the
$e$-foldings number $N=\ln \frac{a}{a_0}$, and so we obtain,
\begin{equation}
\label{RZ4}
0 = - \frac{F(R)}{2} + 3 \left( H^2 + H H'\right) F'(R)
 - 18 (\left(4 H^3 H' + H^2 \left(H'\right)^2 + H^3 H''\right) F''(R)
+ \kappa^2 \rho\, ,
\end{equation}
where $H'\equiv dH/dN$ and $H''\equiv d^2 H/dN^2$. We may consider
the situation that the energy density $\rho$ of the matter perfect
fluids present, is given by a sum of the energy densities of the
perfect fluids with constant equation of state (EoS) $w_i$,
\begin{equation}
\label{RZ6}
\rho=\sum_i \rho_{i0} a^{-3(1+w_i)} = \sum_i \rho_{i0} a_0^{-3(1+w_i)} \e^{-3(1+w_i)N}\, .
\end{equation}
In the FRW Universe, the Hubble rate $H$ can be expressed in terms
of the $e$-foldings number $N$, $H=g(N)$ and the scalar curvature
has the form, $R = 6 g'(N) g(N) + 12 g(N)^2$, which could be
solved (if the equation can be inverted) with respect to $N$ as
$N=N(R)$. Then we may rewrite Eq.~(\ref{RZ4}) in terms of the
$e$-foldings number $N$,
\begin{align}
\label{RZ9}
0 =& -18 \left(4g\left(N\left(R\right)\right)^3 g'\left(N\left(R\right)\right)
+ g\left(N\left(R\right)\right)^2 g'\left(N\left(R\right)\right)^2
+ g\left(N\left(R\right)\right)^3g''\left(N\left(R\right)\right)\right) \frac{d^2 F(R)}{dR^2} \nn
& + 3 \left( g\left(N\left(R\right)\right)^2
+ g'\left(N\left(R\right)\right) g\left(N\left(R\right)\right)\right) \frac{dF(R)}{dR}
 - \frac{F(R)}{2}
+ \sum_i \rho_{i0} a_0^{-3(1+w_i)} \e^{-3(1+w_i)N(R)}\, ,
\end{align}
which constitutes a differential equation for the $F(R)$ gravity,
where the variable is the scalar curvature $R$. The expression
(\ref{RZ9}) can be simplified by using $G(N) \equiv
g\left(N\right)^2 = H^2$, so we get,
\begin{align}
\label{RZ11} 0 =& -9 G\left(N\left(R\right)\right)\left(4
G'\left(N\left(R\right)\right) +
G''\left(N\left(R\right)\right)\right) \frac{d^2 F(R)}{dR^2} +
\left( 3 G\left(N\left(R\right)\right) + \frac{3}{2}
G'\left(N\left(R\right)\right) \right) \frac{dF(R)}{dR} \nn & -
\frac{F(R)}{2} + \sum_i \rho_{i0} a_0^{-3(1+w_i)}
\e^{-3(1+w_i)N(R)}\,,
\end{align}
and also $R= 3 G'(N) + 12 G(N)$. Note that given the form of the
Hubble rate, one may solve the above equation (\ref{RZ11}) and can
find directly the form of the $F(R)$ gravity that can realize the
given Hubble rate. Before going to the case of realizing the early
dark energy era, let us exemplify the method by realizing the
$\Lambda$CDM model with $F(R)$ gravity. The $\Lambda$CDM model
Hubble rate has the following form (ignoring for the moment the
radiation),
\begin{equation}
\label{RZ13} \frac{3}{\kappa^2} H^2 = \frac{3}{\kappa^2} H_0^2 +
\rho_0 a^{-3} = \frac{3}{\kappa^2} H_0^2 + \rho_\mathrm{m}^{(0)} a_0^{-3}
\e^{-3N} \, ,
\end{equation}
where $H_0$ is the Hubble constant at present time, $\rho_\mathrm{m}^{(0)}$
is the energy density of cold dark matter (CDM) at present time,
and $a_0$ is the scale factor at present time. The second term in
the right hand side of Eq.~(\ref{RZ13}) corresponds to the
cosmological constant and the third term to the CDM. The (effective)
cosmological constant $\Lambda$ in the present Universe is given
by $\Lambda = 12 H_0^2$. Then one gets,
\begin{equation}
\label{RZ14} G(N) = H_0^2 + \frac{\kappa^2}{3} \rho_\mathrm{m}^{(0)}
a_0^{-3} \e^{-3N} \, ,
\end{equation}
and $R = 3 G'(N) + 12 G(N) = 12 H_0^2 + \kappa^2 \rho_\mathrm{m}^{(0)}
a_0^{-3} \e^{-3N}$, which can be solved with respect to $N$ as
follows,
\begin{equation}
\label{RZ16} N = - \frac{1}{3}\ln \left(\frac{ \left(R - 12
H_0^2\right)}{\kappa^2 \rho_\mathrm{m}^{(0)} a_0^{-3}}\right)\, .
\end{equation}
Then Eq.~(\ref{RZ11}) takes the following form:
\begin{equation}
\label{RZ17}
0=3\left(R - 9H_0^2\right)\left(R - 12H_0^2\right) \frac{d^2 F(R)}{d^2 R}
 - \left( \frac{1}{2} R - 9 H_0^2 \right) \frac{d F(R)}{dR} - \frac{1}{2} F(R)\, .
\end{equation}
By changing the variable from $R$ to $x$ by $x=\frac{R}{3H_0^2} - 3$,
Eq.~(\ref{RZ17}) reduces to the hypergeometric differential equation,
\begin{equation}
\label{RZ19}
0=x(1-x)\frac{d^2 F}{dx^2} + \left(\gamma - \left(\alpha + \beta
+ 1\right)x\right)\frac{dF}{dx}
 - \alpha \beta F\, .
\end{equation}
Here
\begin{equation}
\label{RZ20}
\gamma = - \frac{1}{2}\ ,\alpha + \beta = - \frac{1}{6}\ ,\quad \alpha\beta
= - \frac{1}{6}\, .
\end{equation}
The analytic solution of Eq.~(\ref{RZ19}) is the Gauss
hypergeometric function $F(\alpha,\beta,\gamma;x)$,
\begin{equation}
\label{RZ22} F(x) = A F(\alpha,\beta,\gamma;x) + B x^{1-\gamma}
F(\alpha - \gamma + 1, \beta - \gamma + 1, 2-\gamma;x)\, ,
\end{equation}
where $A$ and $B$ are constant. Thus, we demonstrated that $F(R)$
gravity may realize the $\Lambda$CDM epoch without the need to
introduce an effective cosmological constant. The attribute of
$F(R)$ gravity is that the dark energy era has not a constant
equation of state parameter, but it is dynamical.
In \cite{Doran:2006kp}, a parametrization of the fractional energy
density for the early dark energy was introduced,
\begin{equation}
\Omega_\mathrm{de}(a)= \frac{\Omega_\mathrm{de}^0
 - \Omega_\mathrm{e} \left(1-a^{-3w_0} \right)}{\Omega_\mathrm{de}^0
+\Omega_\mathrm{m}^0 a^{3w_0}} + \Omega_\mathrm{e}
\left(1-a^{-3w_0} \right)\, ,
\end{equation}
where $\Omega_\mathrm{de}^0+\Omega_\mathrm{m}^0\sim 1$ and we
choose $\Omega_\mathrm{e}=0.01$, $\Omega_\mathrm{m}^0 \sim 0.3$,
and $w_0 \sim -1$ (in a later section we shall use more accurate
values for the matter densities). Then the Hubble rate is given by
\begin{equation}
\frac{H^2 (a)}{H_0^2} = \frac{\Omega_\mathrm{m}^0 a^{-3}
+\Omega_\mathrm{rel}^0a^{-4}}{1-\Omega_\mathrm{de}(a)}\,\,,
\end{equation}
where $\Omega_\mathrm{rel}^0$ is fractional energy density of
relativistic particles at present day, with
$\Omega_\mathrm{rel}^0\sim 5\times 10^{-5}$. Then expressing
everything in terms of the $e$-foldings, we get,
\begin{align}
\label{G1}
G(N) =& H(N)^2 = H_0^2 \frac{\Omega_\mathrm{m}^0 \e^{-3N}
+\Omega_\mathrm{rel}^0 \e^{-4N}}{1
 - \frac{\Omega_\mathrm{de}^0
 - \Omega_\mathrm{e} \left(1- \e^{3N} \right)}{\Omega_\mathrm{de}^0
+\Omega_\mathrm{m}^0 \e^{-3N}}
+ \Omega_\mathrm{e} \left(1- \e^{3N} \right)} \nn
=& H_0^2 \frac{\left( \Omega_\mathrm{m}^0 \e^{-3N}
+\Omega_\mathrm{rel}^0 \e^{-4N}\right)
\left( \Omega_\mathrm{de}^0 + \Omega_\mathrm{m}^0 \e^{-3N} \right) }
{\Omega_\mathrm{m}^0 \e^{-3N} + \Omega_\mathrm{e} \left(1- \e^{3N} \right)
+ \Omega_\mathrm{e} \left(1- \e^{3N} \right)
\left( \Omega_\mathrm{de}^0 +\Omega_\mathrm{m}^0 \e^{-3N} \right) } \, .
\end{align}
Here we have put $w_0=-1$ and $a_0=1$, which means that $N=0$ at
present time. Then we find,
\begin{align}
\label{G2}
G'(N) = H_0^2 & \left\{
\frac{\left( -3 \Omega_\mathrm{m}^0 \e^{-3N}
 - 4 \Omega_\mathrm{rel}^0 \e^{-4N}\right)
\left( \Omega_\mathrm{de}^0 + \Omega_\mathrm{m}^0 \e^{-3N} \right)
 - 3 \left( \Omega_\mathrm{m}^0 \e^{-3N} +\Omega_\mathrm{rel}^0 \e^{-4N}\right)
\Omega_\mathrm{m}^0 \e^{-3N} }
{\Omega_\mathrm{m}^0 \e^{-3N} + \Omega_\mathrm{e} \left(1- \e^{3N} \right)
+ \Omega_\mathrm{e} \left(1- \e^{3N} \right)
\left( \Omega_\mathrm{de}^0 +\Omega_\mathrm{m}^0 \e^{-3N} \right) } \right. \nn
& \left. - \frac{\left( \Omega_\mathrm{m}^0 \e^{-3N}
+\Omega_\mathrm{rel}^0 \e^{-4N}\right)
\left( \Omega_\mathrm{de}^0 + \Omega_\mathrm{m}^0 \e^{-3N} \right)
\left( -3 \Omega_\mathrm{m}^0 \e^{-3N} - 3 \Omega_\mathrm{e} \e^{3N}
 - 3 \Omega_\mathrm{e} \left( \Omega_\mathrm{m}^0 \e^{-3N}
+ \Omega_\mathrm{de}^0 \e^{3N} \right) \right) }
{ \left( \Omega_\mathrm{m}^0 \e^{-3N} + \Omega_\mathrm{e} \left(1- \e^{3N} \right)
+ \Omega_\mathrm{e} \left(1- \e^{3N} \right)
\left( \Omega_\mathrm{de}^0 +\Omega_\mathrm{m}^0 \e^{-3N} \right)\right)^2 }
\right\} \, .
\end{align}
Then after solving the algebraic equation $R = 6 g'(N) g(N) + 12
g(N)^2$ with respect to $N$ as $N=N(R)$, we may solve the
differential equation (\ref{RZ11}), which gives the functional
form of the $F(R)$ implicitly. However, it is rather difficult to
extract analytic results from the above differential equation.
Therefore, in the next section we shall adopt a more qualitative
approach in order to provide a model for early dark energy, and
simultaneously the same model can describe the late-time era too.
 We also discuss several models of early and late-time dark
energy, and we quantify our qualitative proposals by numerically
integrating the Friedman equation.

We also need to note that in principle it is possible to describe
a smooth transition between several cosmic eras, by appropriately
choosing the Hubble rate. This Hubble rate choice can in principle
also describe any possible transitions between the various eras,
however, an analytic treatment of this case might be hard. The
interested reader may consult Ref. \cite{Odintsov:2016plw}, were
this issue is analytically addressed.

\section{Realization of Early Dark Energy with $F(R)$ Gravity: A Qualitative Approach}

Since we showed that it is rather difficult to construct the
$F(R)$ gravity model that may realize the early dark energy
quantitatively, we now try to construct a model by adopting a
qualitative approach. We may consider the following type of $F(R)$
gravity,
\begin{equation}
\label{triFR1}
F(R) = R + F_\mathrm{inf} (R) + F_\mathrm{EDE} (R)
+ F_\mathrm{DE} (R) \, ,
\end{equation}
where the term $F_\mathrm{inf} (R)$ is chosen in such a way so
that it dominates in the high curvature regime, so during the
inflationary era. Also the term $F_\mathrm{DE} (R)$ can be chosen
in such a way so that it dominates in the late-time era, and can
describe the dark energy fluid. For example, we may choose
\begin{equation}
\label{triFR2}
F_\mathrm{inf} (R) = \frac{R^2}{M^2} \, , \quad
F_\mathrm{DE} (R) = 2 \Lambda \left( \e^{- \frac{R}{R_l}} - 1
\right) \, ,
\end{equation}
but there are also several other examples which may realize
successfully an early and a late-time era. Another interesting
model firstly appeared in the end of \cite{Odintsov:2019evb}, in
which case the functional form of the $F(R)$ gravity is,
\begin{equation}
\label{starobinsky1}
R + F_\mathrm{inf} (R) + F_\mathrm{DE} (R)
=R + \frac{R^2}{M^2} - \gamma \Lambda
\left( \frac{R}{3m_s^2} \right)^{\delta}\, ,
\end{equation}
where $\delta$ is a positive number $0<\delta <1$ and the rest of
the parameters will be defined later on in this section. In the
model of Eq.~(\ref{starobinsky1}) the $R^2$ term drives the
inflationary era, while the term $\sim R^{\delta}$ drives the
late-time era.

Returning our focus on Eq.~(\ref{triFR1}), the term
$F_\mathrm{EDE} (R)$ can be chosen in such a way so that it may
realize the early dark energy era. In order to construct a model
of $F_\mathrm{EDE} (R)$, we should recall the following
\cite{Cognola:2007zu,Nojiri:2010wj} issues: When $F(R)$ is written
in the neighborhood of $R\sim R_0$ as,
\begin{equation}
\label{FRp1}
\frac{F(R)}{R^2} \sim f_0 + f(R) \left( R - R_0 \right)^n \, ,
\end{equation}
with constants $f_0>0$, $R_0>0$, and a positive integer $n$, if a
function $f(R)$ does not vanish at $R=R_0$, $f\left( R_0
\right)\neq 0$, and $R=R_0$ is an exact solution describing the de
Sitter space-time. It has been shown that
\begin{enumerate}
\item\label{i1} When $n$ is an even integer and $f\left( R_0
\right)<0$, the solution describing the de Sitter space-time is
stable. \item\label{i2} When $n$ is an even integer and $f\left(
R_0 \right)>0$, the solution describing the de Sitter space-time
is unstable. \item\label{i3} When $n$ is an odd integer, the
solution describing the de Sitter space-time is quasi-stable. If
$f\left( R_0 \right)<0$, the curvature $R$ decreases by crossing
$R=R_0$. \item\label{i4} When $n$ is an odd integer, the solution
describing the de Sitter space-time is quasi-stable. If $f\left(
R_0 \right)>0$, the curvature $R$ increases by crossing $R=R_0$.
\end{enumerate}
The third case may describe the early dark energy since the
quasi-stable de Sitter solution would realize a post-inflationary
early dark energy era. Then for example, we may choose,
\begin{equation}
\label{triFR3}
f(R) = - \frac{\beta R^{m-n-2}}{R_0^{l+m} + R^{l+m}} \, ,
\end{equation}
with a constant $\beta>0$ of dimensions $[m]^{-l}$, and we can
choose the integers $m>0$ and $l>0$ to be large enough. We also
choose $R_0$ in such a way so that $R_0$ is slightly larger than
the curvature of the Universe during the recombination epoch. Then
we find,
\begin{equation}
\label{triFR4}
F_\mathrm{EDE} (R) = f_0 R^2
 - \frac{\beta R^{m-n} \left( R - R_0 \right)^n}{R_0^{l+m} + R^{l+m}} \, .
\end{equation}
The first term in (\ref{triFR4}) can be absorbed into the
redefinition of $\frac{1}{M^2}$ in $F_\mathrm{inf} (R)$ if we
choose $F_\mathrm{inf} (R)$ as in (\ref{triFR2}) and therefore we
put $f_0=0$ in the following. We should note that the second term
in Eq.~(\ref{triFR4}) decreases in the early Universe where $R$ is
large and in the late Universe where $R$ is small because $m$ and
$l$ are large enough. In the early Universe, where the curvature
$R$ could be large and we assume $R\gg R_0$, if we choose
$\Lambda$ in (\ref{triFR2}) small enough and $l$ in (\ref{triFR3})
or (\ref{triFR3}) large enough, so that,
\begin{equation}
\label{triFR5} \left| F_\mathrm{EDE} (R) \right| \sim
\left|\frac{\beta}{R^l}\right| \ll \left| F_\mathrm{inf} (R)
\right| = \left| \frac{1}{M^2} R^2 \right|\, ,
\end{equation}
the second term $F_\mathrm{inf} (R)$ in (\ref{triFR1}) becomes
dominant and generates the inflationary era. When the curvature
becomes smaller, $R\sim R_0$, if we choose $\left|\frac{1}{M^2}
R_0^2\right| \ll \beta R_0^{-l}$ and $m$ is large enough, the
third term $F_\mathrm{EDE} (R)$ could dominate except right on the
point $R=R_0$ and generate the quasi-de Sitter space-time, which
may be identified with the early dark energy era. Since the de
Sitter space-time solution at the point $R=R_0$ is quasi-stable,
the curvature becomes smaller very slowly and after that the
recombination era occurs. After the recombination era, the
curvature $R$ decreases significantly and eventually, since we
chose $m$ to be large enough, the fourth term $F_\mathrm{DE} (R)$
in (\ref{triFR1}) becomes much larger than the third term
$F_\mathrm{EDE} (R)$,
\begin{equation}
\label{triFR6}
\left| F_\mathrm{EDE} (R) \right| \sim \left|\frac{\beta R^{m-n}}{R_0^{l+m-n}}\right|
\ll \left| F_\mathrm{DE} (R) \right| \sim 2 \left| \Lambda \right|\, .
\end{equation}
We also choose the parameter $M$ of $F_\mathrm{inf} (R)$ in
(\ref{triFR2}), in such a way so that the fourth term
$F_\mathrm{DE} (R)$ in (\ref{triFR1}) becomes much larger than the
second term $F_\mathrm{inf} (R)$, that is,
\begin{equation}
\label{triFR7} \left| F_\mathrm{inf} (R) \right| = \frac{1}{M^2}
\left| R^2 \right| \ll \left| F_\mathrm{DE} (R) \right| \sim 2
\left| \Lambda \right|\, .
\end{equation}
when $R\ll R_0$. When $R_0\gg R \gg R_l$ as in the Universe at
present time, the fourth term $F_\mathrm{DE} (R)$ in
(\ref{triFR1}) becomes dominant and behaves as a constant
$F_\mathrm{DE} (R) \sim 2 \Lambda$, which plays the role of the
effective cosmological constant in the present Universe. Then by
adjusting the parameters in $F_\mathrm{EDE} (R)$, we could in
principle modify the Hubble constant in the present Universe, in
such a way so that the Hubble constant is not conflict with the
value given by the observations of the Cepheid variables. Since
$F_\mathrm{EDE} (R)$ affects strongly in a decreasing way both the
the large curvature and the small curvature regimes, the term does
not influence the inflationary era, where the curvature is large,
and also the accelerating expansion of the present Universe, where
the curvature is small. Another model unifying the inflationary
and the late-acceleration eras has been proposed in
\cite{Elizalde:2010ts},
\begin{equation}
\label{total}
R + F_\mathrm{inf} (R) + F_\mathrm{DE} (R)
=R-2\Lambda_l\left(1-\mathrm{e}^{-\frac{R}{R_{l}}}\right)
-\Lambda_{I}\left(1-\mathrm{e}^{-\left(\frac{R}{R_I}\right)^n}\right)
+\gamma R^\alpha\, .
\end{equation}
The last term with constant $\gamma$ and $1<\alpha\leq 2$ are
added in order to avoid that the curvature singularity appears in
the dense matter regions (see \cite{Cognola:2007zu}, for example).
The constants $R_l$ and $R_I$ correspond to the curvature in the
inflationary epoch and in the late-time acceleration epoch,
respectively, and $\Lambda_l$ and $\Lambda_I$ are constants which
act as effective cosmological constants during the inflationary
and the late-time acceleration epochs, respectively. Then by
adding $F_\mathrm{EDE} (R)$ to (\ref{total}), we may also realize
the early dark energy.

Now let us analyze the effects of $F_\mathrm{EDE} (R)$ during the
late-time epoch in a more concrete way, and see at first hand the
range of values of the free parameters for which the effect of
$F_\mathrm{EDE} (R)$ on the late-time era is insignificant. We
shall analyze the model (\ref{starobinsky1}) with the addition of
$F_\mathrm{EDE} (R)$, so that the final $F(R)$ gravity is,
\begin{equation}
\label{starobinsky}
F(R)=R+\frac{R^2}{M^2} - \gamma \Lambda
\left( \frac{R}{3m_s^2} \right)^{\delta} - \frac{\beta R^{m-n}
\left( R - R_0 \right)^n}{R_0^{l+m} + R^{l+m}} \, .
\end{equation}
We shall also assume that the Universe is filled with dust and
radiation perfect fluids, for the late-time study of the model
(\ref{starobinsky}). The Universe's evolution is affected by the
geometry via the $F(R)$ gravity, so the geometry contributes an
energy density and a pressure in the Friedmann and Raychaudhuri
equation, which for the FRW space-time is written as follows,
\begin{equation}
\label{flat}
3H^2=\kappa^2\rho_\mathrm{tot}\, ,\quad
 -2\dot{H}=\kappa^2(\rho_\mathrm{tot}+P_\mathrm{tot})\, ,
\end{equation}
with $\rho_\mathrm{tot}=\rho_\mathrm{m}+\rho_\mathrm{G}
+\rho_\mathrm{r}$ being the total energy
density of the Universe, and accordingly the total pressure is
$P_\mathrm{tot}=P_\mathrm{r}+P_\mathrm{G}$. Also $\rho_\mathrm{r}$
is the energy density of relativistic matter,
$P_\mathrm{r}$ is its pressure, and $\rho_\mathrm{G}$ is the energy density of the
geometric fluid,
\begin{equation}
\label{degeometricfluid} \rho_\mathrm{G}=\frac{F_R
R-F}{2}+3H^2(1-F_R)-3H\dot{F}_R\, ,
\end{equation}
where $F_R=\frac{\partial F}{\partial R}$. The geometric fluid
will drive inflation at early times, and eventually will also
produce the early dark energy era, and accordingly will drive the
late-time evolution. The corresponding pressure of the geometric
fluid is equal to,
\begin{equation}
\label{pressuregeometry}
P_\mathrm{G}=\ddot{F}_R-H\dot{F}_R+2\dot{H}(F_R-1)-\rho_\mathrm{G}\, .
\end{equation}
It is easy to show that all the fluids satisfy the energy
conservation conditions,
\begin{equation}
\label{fluidcontinuityequations}
\dot{\rho}_a+3H(\rho_a+P_a)=0\, , \quad
\dot{\rho}_\mathrm{r}+3H(\rho_\mathrm{r}+P_\mathrm{r})=0\, , \quad
\dot{\rho}_\mathrm{G}+3H(\rho_\mathrm{G}+P_\mathrm{G})=0\, .
\end{equation}
Our aim in the rest of this section is to numerically solve the
cosmological equations, by using appropriate variables and
specific values for the free parameters that may yield a viable
late-time phenomenology. We start off by specifying the values of
the free parameters of the model (\ref{starobinsky}), so the
parameter $M$ for phenomenological reasons related to the
inflationary era must be \cite{Appleby:2009uf},
\begin{equation}
\label{M}
M= 1.5\times 10^{-5}\left(\frac{N}{50}\right)^{-1}M_\mathrm{P}\, ,
\end{equation}
where $N$ is the $e$-foldings number during inflation, and $M_\mathrm{P}$
is the reduced Planck mass $M_\mathrm{P}=2.435\times 10^{27}\,\mathrm{eV}$. Hence for
$N\sim 60$, $M$ takes approximately the value $M\simeq
3.04375\times 10^{22}\,\mathrm{eV}$. In addition, $m_s$ in
Eq.~(\ref{starobinsky}) shall be taken equal to
$m_s^2=\frac{\kappa^2\rho_\mathrm{m}^{(0)}}{3}$, and in addition we shall
assume that $\Lambda\simeq 11.895\times 10^{-67}\, \mathrm{eV}^2$. The
latest Planck data on the cosmological parameters indicate that
\cite{Aghanim:2018eyx},
\begin{equation}
\label{H0today}
H_0=67.4\pm 0.5\, \mathrm{km}/\left(\mathrm{sec}\cdot\mathrm{Mpc}\right)\, ,
\end{equation}
hence $H_0=67.4\, \mathrm{km}/\left(\mathrm{sec}\cdot\mathrm{Mpc}\right)$
or equivalently $H_0=1.37187\times
10^{-33}\,\mathrm{eV}$, therefore $h\simeq 0.67$. Moreover, the latest Planck
data also indicate that $\Omega_c h^2$ is,
\begin{equation}\label{codrdarkmatter}
\Omega_c h^2=0.12\pm 0.001\, ,
\end{equation}
in effect, the parameter $m_s^2$ is,
\begin{equation}\label{msquare}
m_s^2=\frac{\kappa^2\rho^{(0)}_\mathrm{m}}{3}=H_0\Omega_c=1.37201\times
10^{-67}\,\mathrm{eV}^2\, .
\end{equation}
\begin{figure}[h!]
\centering
\includegraphics[width=18pc]{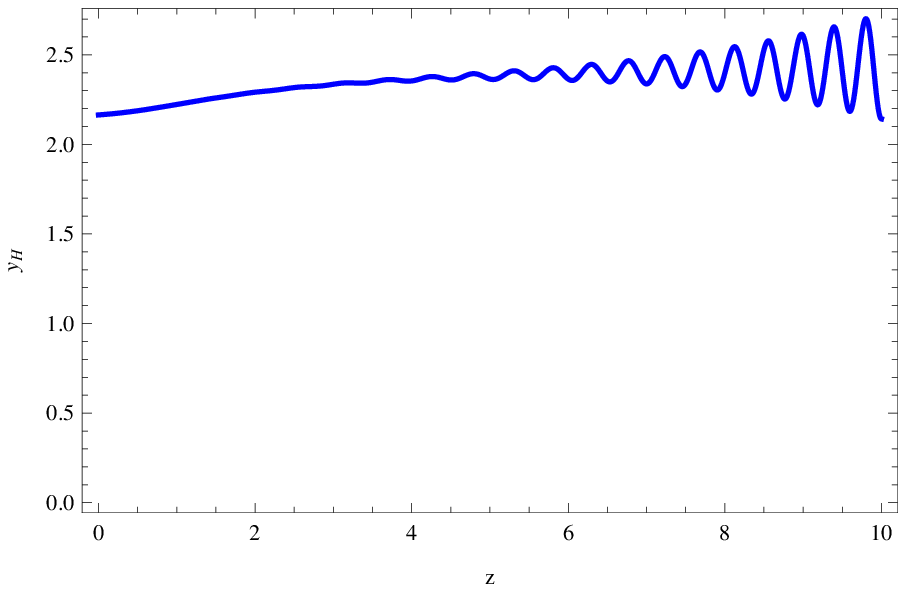}
\includegraphics[width=18pc]{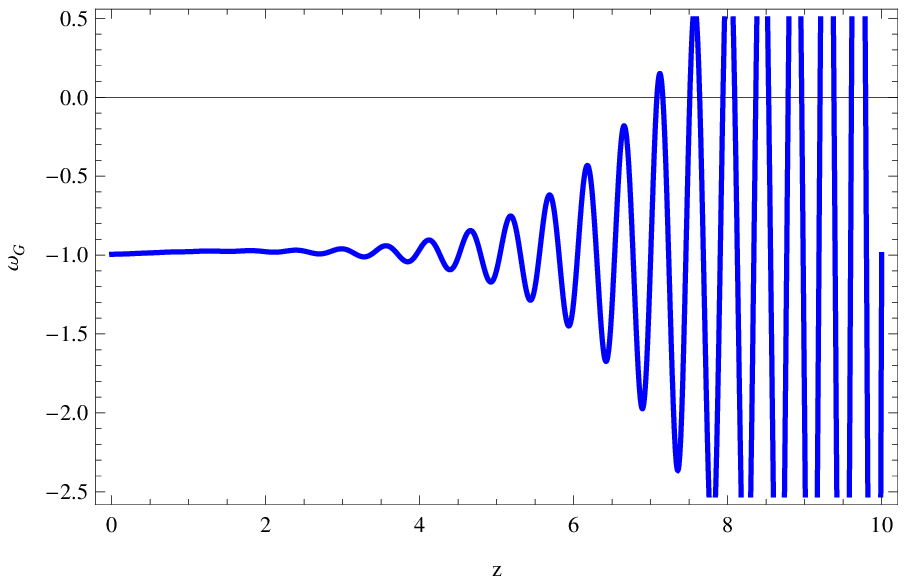}
\includegraphics[width=18pc]{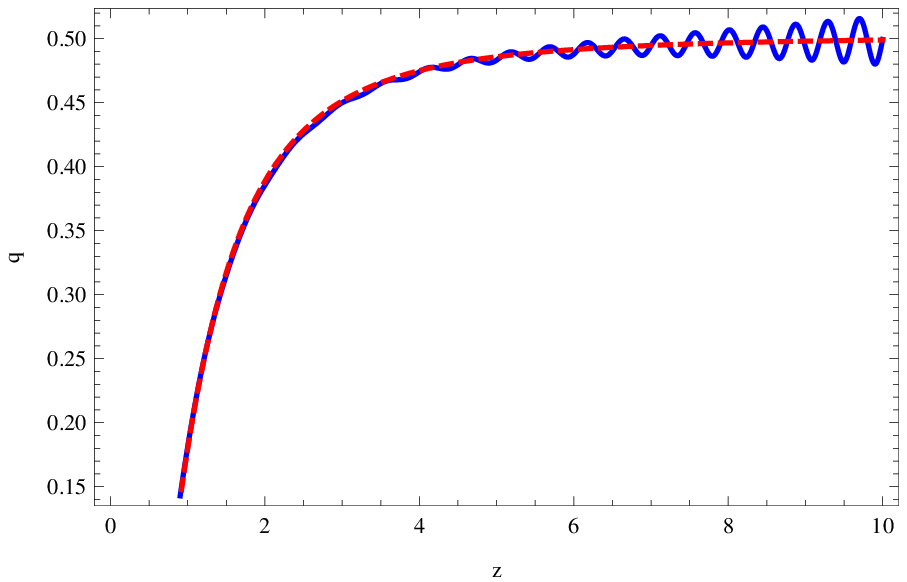}
\caption{The function $y_\mathrm{H}$ (left upper plot), the dark energy EoS
parameter $\omega_\mathrm{G}(z)$ (right upper plot) and the deceleration
parameter (bottom plot), as functions of the redshift, for the
$F(R)$ gravity model without the early dark energy term.}
\label{plot1}
\end{figure}
The parameters $\delta$ and $\gamma$ will be assumed to take the
values $\delta =1/100$ and $\gamma=2$. The phenomenology of the
model (\ref{starobinsky}) without the early dark energy term is
quite interesting and is studied in detail in
\cite{Odintsov:2020nwm}, however we shall briefly show that it
indeed yields a viable phenomenology at late-times. Then we shall
include the early dark energy term in order to investigate when
the early dark energy term affects the late-time era, and then
constrain the parameter $\beta $. For our late-time era analysis,
we shall express all the quantities involved in the cosmological
equations as functions of the redshift $z$,
\begin{equation}
\label{redshift}
1+z=\frac{1}{a}\, ,
\end{equation}
by also taking the scale factor of the current Universe at $z=0$
equal to unity. Moreover, we introduce the statefinder quantity
$y_\mathrm{H}(z)$ \cite{Hu:2007nk,Bamba:2012qi}, which is equal to,
\begin{equation}
\label{yHdefinition}
y_\mathrm{H}(z)=\frac{\rho_\mathrm{G}}{\rho^{(0)}_\mathrm{m}}\, ,
\end{equation}
so by using the first Friedmann equation (\ref{flat}), $y_\mathrm{H}(z)$
takes the following form,
\begin{equation}
\label{yhfunctionanalyticzero}
y_\mathrm{H}(z)=\frac{3H^2}{\kappa^2\rho^{(0)}_\mathrm{m}}
 -\frac{\rho_\mathrm{m}}{\rho^{(0)}_\mathrm{m}}
 -\frac{\rho_\mathrm{r}}{\rho^{(0)}_\mathrm{m}}\, .
\end{equation}
Due to the fact that $\rho_\mathrm{r}=\rho_\mathrm{r}^{(0)}a^{-4}$, which means
$\frac{\rho_\mathrm{r}}{\rho^{(0)}_\mathrm{m}}=\chi (1+z)^4$, with
$\chi=\frac{\rho^{(0)}_\mathrm{r}}{\rho^{(0)}_\mathrm{m}}\simeq 3.1\times 10^{-4}$,
and $\rho_\mathrm{m}=\rho_\mathrm{m}^{(0)}$, $y_\mathrm{H}(z)$ is equal to,
\begin{equation}
\label{finalexpressionyHz}
y_\mathrm{H}(z)=\frac{H^2}{m_s^2}-(1+z)^{3}-\chi (1+z)^4\, .
\end{equation}
We can easily express the Friedmann equations in terms of
$y_\mathrm{H}(z)$, so it reads \cite{Bamba:2012qi},
\begin{equation}
\label{differentialequationmain}
\frac{d^2y_\mathrm{H}(z)}{d z^2}+J_1\frac{d y_\mathrm{H}(z)}{d z}
+J_2y_\mathrm{H}(z)+J_3=0\, ,
\end{equation}
with the functions $J_1$, $J_2$ and $J_3$ being defined in the
following way,
\begin{align}
\label{diffequation}
J_1=&\frac{1}{z+1}\left(
 -3-\frac{1-F_R}{\left(y_\mathrm{H}(z)+(z+1)^3+\chi (1+z)^4\right) 6
m_s^2F_{RR}} \right)\, , \nn
J_2=&\frac{1}{(z+1)^2}\left(
\frac{2-F_R}{\left(y_\mathrm{H}(z)+(z+1)^3+\chi (1+z)^4\right) 3
m_s^2F_{RR}} \right)\, ,\nn
J_3=&-3(z+1)-\frac{\left(1-F_R\right)\left( (z+1)^3+2\chi (1+z)^4
\right) +\frac{R-F}{3m_s^2}}{(1+z)^2
\left( y_\mathrm{H}(z)+(1+z)^3+\chi(1+z)^4\right) 6m_s^2F_{RR}}\, ,
\end{align}
with $F_{RR}=\frac{\partial^2 F}{\partial R^2}$. We shall
numerically solve the differential equation
(\ref{differentialequationmain}), in the redshift interval
$z=[z_i,z_f]=[0,10]$ with the initial conditions for the function
$y_\mathrm{H}(z)$ at redshift $z_f=10$ chosen as follows,
\begin{equation}
\label{generalinitialconditions}
y_\mathrm{H}(z_f)=\frac{\Lambda}{3m_s^2}\left(
1+\frac{1}{1000}(1+z_f)\right)\, , \quad
\left. \frac{d y_\mathrm{H}(z)}{d z}\right|_{z=z_f}=\frac{1}{1000}\frac{\Lambda}{3m_s^2}\, .
\end{equation}
Let us now present the results of the numerical analysis, focusing
first briefly on the model (\ref{starobinsky}) without the
presence of the early dark energy term. The full analysis of this
model is performed elsewhere \cite{Odintsov:2020nwm}.
In order to have some standard comparison reference, we shall
compare the results of the model (\ref{starobinsky}) without the
early dark energy term, with the $\Lambda$CDM model. Accordingly,
we shall compare the model without the early dark energy term,
with the model (\ref{starobinsky}), in the presence of the early
dark energy term, in order to see which values of the parameter
$\beta$ can affect the late-time behavior. Let us start off our
analysis with the phenomenology of the model (\ref{starobinsky})
without the early dark energy term, and by numerically solving the
differential equation (\ref{differentialequationmain}), we present
the solution $y_\mathrm{H}(z)$ in the left plot of
Fig.~\ref{plot1}. In the right plot of Fig.~\ref{plot1}, we
present the dark energy EoS parameter $\omega_\mathrm{G}(z)$, and
in the bottom  plot we present the deceleration parameter $q(z)$.
In all the plots, the blue curves correspond to the $F(R)$ gravity
model (\ref{starobinsky}) without the early dark energy term, and
the red curves correspond to the $\Lambda$CDM model. Also at
$z=0$, the value of the dark energy EoS parameter is evaluated to
be $\omega_\mathrm{G}(0)=-0.995827$, and also the dark energy
density parameter is $\Omega_\mathrm{G}(0)=0.683968$, which are
within the constraints of the latest Planck data on the
cosmological parameters. Moreover, the deceleration parameter is
$q(0)=-0.520954$, which is very close to the $\Lambda$CDM value
$q_{\Lambda}=-0.5$. These results indicate that the model can be
considered as phenomenologically viable.
\begin{figure}[h!]
\centering
\includegraphics[width=18pc]{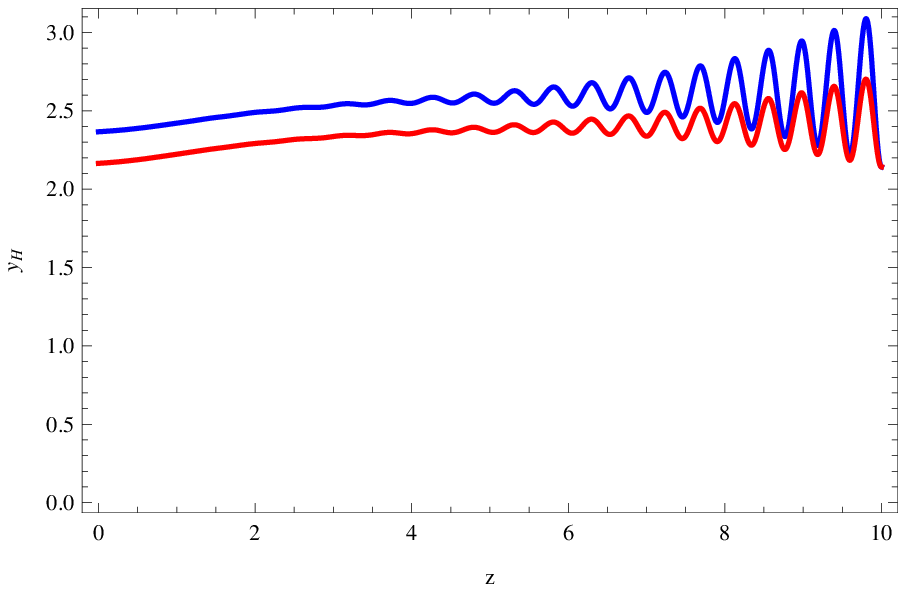}
\includegraphics[width=18pc]{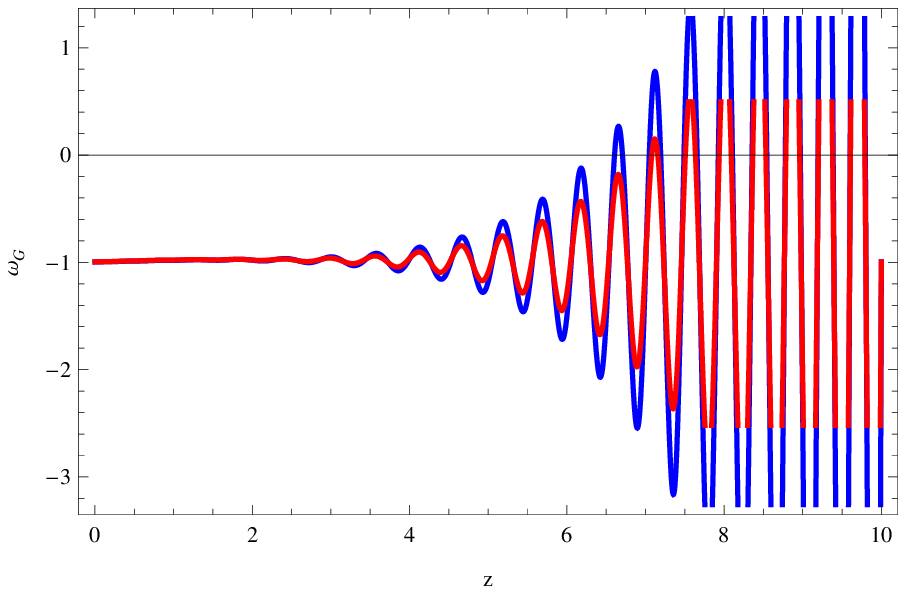}
\includegraphics[width=18pc]{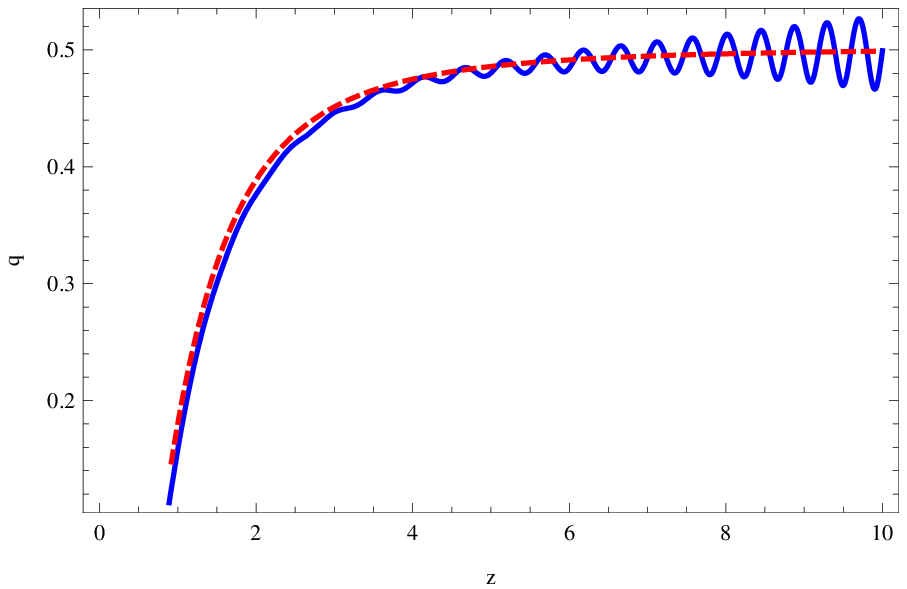}
\caption{The function $y_\mathrm{H}$ (left upper plot), the dark energy EoS
parameter $\omega_\mathrm{G}(z)$ (right upper plot) and the deceleration
parameter (bottom plot), as functions of the redshift, for the
$F(R)$ gravity model with the early dark energy term (blue
curves), for $\beta\geq 10^{400}R_0^l$. The red curves correspond
to the model without the early dark energy term.} \label{plot2}
\end{figure}
Now let us turn our focus on the model (\ref{starobinsky}), by
taking into account the presence of the early dark energy term.
Since the early dark energy era is engineered to occur around the
recombination era, so for $z\sim 1100$, the curvature of the
Universe $R_0$ at this era is of the order
$\mathcal{O}(10^{-59})\,\mathrm{eV}^2$, we shall assume that
$R_0=10^{-59}\,\mathrm{eV}^2$. Also $n$ must be an odd integer in order for
the early dark energy era to have an unstable de Sitter attractor,
hence we take $n=3$. Furthermore, $l$ and $m$ must be larger from
$n$, so we choose $l=m=8$. As it seems, the early dark energy term
has no effect on the late-time era, when $\beta \sim R_0^l$,
however, when $\beta\geq 10^{400}R_0^l$, the early dark energy
term starts to affect the late-time era. Indeed this can be seen
in Fig.~\ref{plot2}, where we plot the statefinder $y_\mathrm{H}(z)$ (left
upper plot), the dark energy EoS parameter (right upper plot) and
the deceleration parameter $q(z)$ (bottom plot). In all the plots
of Fig.~\ref{plot2}, the blue curves correspond to the model
(\ref{starobinsky}), while the red curves to the model
(\ref{starobinsky}) without the early dark energy term. As it can
be seen, the effects of the early dark energy term on the
late-time evolution are significant, but this occurs only when
$\beta\geq 10^{400}R_0^l$. Also the value of the dark energy EoS
parameter is $\omega_\mathrm{G}(0)=-0.996237$, the dark energy density
parameter $\Omega_\mathrm{G}(0)=0.702882$ and the deceleration parameter
$q(0)=-0.55031$, so the effect of the early dark energy term
affect the late-time dynamics for $\beta\geq 10^{400}R_0^l$.

\begin{figure}[h]
\centering
\includegraphics[width=18pc]{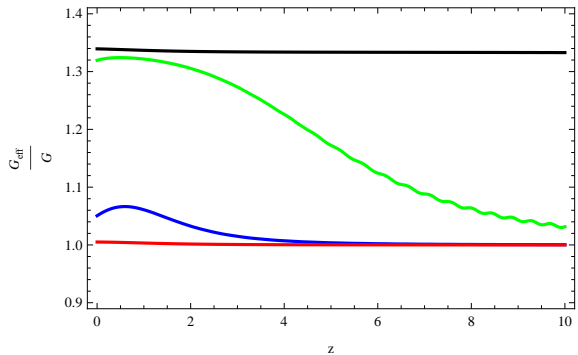}
\caption{Plots of $\frac{G_\mathrm{eff}}{G}$ as a function of $z$, for
$k=1\, \mathrm{Mpc}^{-1}$ (black curve), $k=0.0125\, \mathrm{Mpc}^{-1}$ (green curve),
$k=0.00125\, \mathrm{Mpc}^{-1}$ (blue curve) and $k=0.000125\, \mathrm{Mpc}^{-1}$ (red
curve).}\label{plot3}
\end{figure}

\begin{figure}[h]
\centering
\includegraphics[width=18pc]{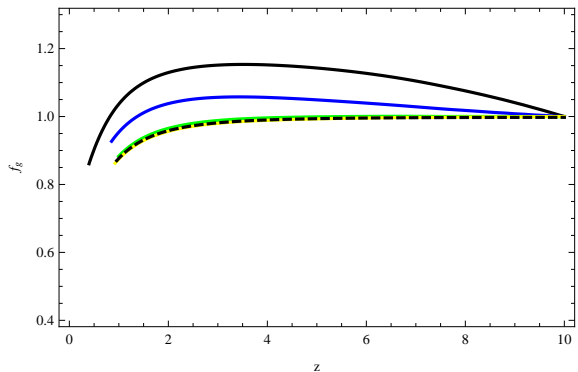}
\caption{Plots of the growth factor $f_\mathrm{g}(z)=\frac{\mathrm{d}\ln
\delta}{\mathrm{d}\ln a}$, as a function of the redshift $z$,
$k=1\, \mathrm{Mpc}^{-1}$ (black curve), $k=0.0125\, \mathrm{Mpc}^{-1}$ (blue curve),
$k=0.00125\, \mathrm{Mpc}^{-1}$ (green curve) and $k=0.000125\, \mathrm{Mpc}^{-1}$
(yellow curve), for the model (\ref{starobinsky1}). The dashed
black curve corresponds to the $\Lambda$CDM model.}\label{plot4}
\end{figure}

Another important issue to discuss in the matter density
perturbations issue and the corresponding growth factor. In fact,
the matter density perturbations can be considered as a consistent
criterion that can distinguish the cosmic evolution corresponding
to different $F(R)$ gravities. This is because the cosmological
matter density perturbations actually make possible the
distinction between the evolution of each $F(R)$ gravity from the
background \cite{Matsumoto:2011ne}. With regard to the matter
density perturbations, a consistent calculation of these requires
the subhorizon approximation, so that the theoretical framework is
consistent with the Newtonian gravity \cite{Matsumoto:2011ne}. The
subhorizon approximation is quantified by the condition,
\begin{equation}
\label{subhorapprx}
\frac{k^2}{a^2}\gg H^2\, ,
\end{equation}
where $a(t)$ is the scale factor and $k$ is the wavenumber of the
comoving mode. Practically, the subhorizon approximation indicates
that we should consider comoving wavelengths $\lambda =a/k$ of a
comoving mode $k$, which are much shorter than the corresponding
Hubble radius $H^{-1}$. The subhorizon approximation holds true
during the matter domination era, and we shall focus on the last
stage of the matter domination era and the dawn of the dark energy
era, so for redshifts $z=[0,10]$. The matter density perturbations
are quantified by the parameter $\delta =\frac{\delta
\rho_\mathrm{m}}{\rho_\mathrm{m}}$, which satisfies the following
differential equation \cite{Bamba:2012qi},
\begin{equation}
\label{matterperturb}
\ddot{\delta}+2H\dot{\delta}-4\pi G_\mathrm{eff}(a,k)\rho_\mathrm{m}\delta =0
\end{equation}
where $G_\mathrm{eff}(a,k)$ is the effective gravitational constant of
the $F(R)$ gravity theory, with its analytic form being for $F(R)$
gravity \cite{Bamba:2012qi},
\begin{equation}
\label{geff}
G_\mathrm{eff}(a,k)=\frac{G}{F'(R)}
\left[1+\frac{\frac{k^2}{a^2}\frac{F''(R)}{F'(R)}}{1+3\frac{k^2}{a^2}\frac{F''(R)}{F'(R)}}
\right] \, ,
\end{equation}
and $G$ denotes the present value of Newton's constant of gravity.
In the following we shall focus on the growth factor
$f_\mathrm{g}(z)=\frac{\mathrm{d}\ln \delta}{\mathrm{d}\ln a}$, so we can
express the differential equation governing the matter density
perturbations evolution (\ref{matterperturb}) in terms of the
growth factor. By using the following rules,
\begin{equation}
\label{basicrelations}
\dot{H}=\frac{\mathrm{d}H}{\mathrm{d}z}(z+1)H(z)\, , \quad
\dot{\delta}=H\dot{f_\mathrm{g}}\delta\, , \quad
\ddot{\delta}=\dot{H}f_\mathrm{g}\delta+H\dot{f_\mathrm{g}}\delta
+Hf_\mathrm{g}\dot{\delta}\, ,
\end{equation}
we can express the differential equation that governs the
evolution of the matter perturbations (\ref{matterperturb}), in
terms of the growth factor,
\begin{equation}
\label{presenceofcoll}
\frac{\mathrm{d}f_\mathrm{g}(z)}{\mathrm{d}z}
+\left( \frac{1+z}{H(z)}\frac{\mathrm{d}H(z)}{\mathrm{d}z}-2
 -f_\mathrm{g}(z)\right) \frac{f_\mathrm{g}(z)}{1+z}
+\frac{4\pi}{G}\frac{G_\mathrm{eff}(a(z),k)}{(z+1)H(z)^2}\rho_\mathrm{m}=0\, .
\end{equation}
 From the above
equation it is apparent that the effect of $F(R)$ gravity on the
growth factor is quantified in the term
$G_\mathrm{eff}\left(a(z),k\right)$, and it is notable that the
latter depends on the comoving wavenumber $k$. By taking the
present time Universe $a=1$, the comoving number can be
constrained by the present time value of the Hubble rate, so since
$k\gg H_0$, the wavenumber $k$ must be $k>0.000124011 \,
\mathrm{Mpc}^{-1}$. In the following we shall numerically solve
the differential equation (\ref{presenceofcoll}), and we shall
investigate two main issues, firstly the general behavior of the
growth factor as a function of the redshift for various allowed
values of the comoving wavenumber $k$, for the model
(\ref{starobinsky1}). Accordingly, we shall consider the model
(\ref{starobinsky}) in which the early dark energy term is added,
and we shall investigate the behavior of the model and examine
whether the early dark energy term affects the late-time behavior
of the growth factor. This study shall validate the assumptions
made in the previous late-time study, which indicated that when
$\beta\sim R_0^l$, the early dark energy term has no effect on the
late-time evolution. For the numerical study, we shall assume that
the growth factor at a redshift $z_f=10$ is equal to,
$f_\mathrm{g}(z_f, k)=0.997595$ \cite{Bamba:2012qi}, which is the
value of the growth factor for the $\Lambda$CDM model. For
comparison reasons, we shall also compare the $F(R)$ gravity model
(\ref{starobinsky1}) without the early dark energy term with the
$\Lambda$CDM, so at this point we quote the $\Lambda$CDM
expression for the growth factor, which is
\cite{Basilakos:2012ws},
\begin{equation}
\label{LCDMgrowthfactor} f_\mathrm{g}(z)=\left(
\frac{H_0^2\Omega_\mathrm{m}^{(0)}(1+z)^3}{H(z)^2}\right)^{\frac{6}{11}}
\, ,
\end{equation}
where $H(z)$ in this case is the $\Lambda$CDM model Hubble rate.
Our aim at this point is to firstly indicate explicitly that the
$F(R)$ gravity model of Eq.~(\ref{starobinsky1}) behaves similarly
to the $\Lambda$CDM model, and secondly, to investigate if the
addition of the early dark energy era term affects the late-time
behavior, and to which extent, having in mind that when $\beta
\sim R_0^l$ the early dark energy term does not affect the
late-time era, as we saw in the previous section. Let us first
focus on the model (\ref{starobinsky1}) and in Fig.~\ref{plot3} we
present the behavior of the fraction $\frac{G_\mathrm{eff}}{G}$
for $k=1\, \mathrm{Mpc}^{-1}$ (black curve), $k=0.0125\,
\mathrm{Mpc}^{-1}$ (green curve), $k=0.00125\, \mathrm{Mpc}^{-1}$
(blue curve) and $k=0.000125\, \mathrm{Mpc}^{-1}$ (red curve). As
it can be seen, as the limiting allowed $k$ value is reached (red
and blue curves), the effective gravitational constant approaches
the present day Newton's value.
Also in Fig.~\ref{plot4} we present the results of the numerical
solution to the equation (\ref{presenceofcoll}) which yields the
growth factor $f_\mathrm{g}(z)$ as a function of the redshift, for
$k=1\, \mathrm{Mpc}^{-1}$ (black curve), $k=0.0125\, \mathrm{Mpc}^{-1}$ (blue curve),
$k=0.00125\, \mathrm{Mpc}^{-1}$ (green curve) and $k=0.000125\, \mathrm{Mpc}^{-1}$
(yellow curve). Also the black dashed curve corresponds to the
growth factor of the $\Lambda$CDM model. As it can be seen, the
green and yellow curves are indistinguishable from the
$\Lambda$CDM model, and deviations occur only for higher values of
the wavenumber.
The results of the numerical analysis hold true, even when the
early dark energy term is added, hence even for the model
(\ref{starobinsky}) and remarkably, no deviations occur even when
$\beta\geq 10^{400}R_0^l$. Therefore in conclusion, the dark
energy era of the model (\ref{starobinsky}) is not affected at all
from the early dark energy term. Using the same approach as in
this section, with similar a form of early dark energy, one can
show that the realistic $F(R)$ gravities unifying inflation with
dark energy of Refs.
\cite{Nojiri:2017ncd,Nojiri:2007as,Nojiri:2007cq,Cognola:2007zu,Nojiri:2006gh,Appleby:2007vb,Elizalde:2010ts,Odintsov:2020nwm},
may be easily extended by adding an early dark energy term which
does not spoil the inflationary and dark energy behavior of the
models. For relevant works in the same unification of inflation
with dark energy line of research see for example
\cite{Benisty:2017lmt,Banerjee:2019kgu,Benisty:2019jqz}.

A natural question that springs to mind is whether it is possible
to obtain any new viable condition of $F(R)$ gravity as an
alternative gravity theory to general relativity? In principle, in
the numerical study mainly, we investigated the constraints on the
model (\ref{starobinsky}) imposed by the requirement of the
realization of the early dark energy era, in conjunction with the
requirement that a viable late-time era is realized by the same
model. As we showed, these two requirements severely constrained
the values of the parameters $l$, $m$ and $R_0$.

At this point, let us comment on the two approaches we used in
order to realize an early dark energy era. In the first approach,
we used a rather quantitative method, which can provide the exact
form of $F(R)$ which reproduces an appropriately chosen
phenomenological functional form of the Hubble rate. In this way,
we can find the functional form of $F(R)$ gravity, which
reproduces a particular behavior of the Hubble rate $H$ in
specific era, like matter dominated era, radiation dominated era,
or dark energy dominated era and so on. In the second approach,
however, we do not specify the Hubble rate, so it is analytically
difficult to solve the equations of motion for a specifically
chosen $F(R)$ gravity. To be more specific in the first approach,
the functional form of the Hubble rate is given, and one seeks the
exact form of the $F(R)$ gravity which may realize the
appropriately chosen phenomenological Hubble rate. In the second
approach, the phenomenology of various $F(R)$ gravities containing
terms that may realize an early dark energy is checked, but in
this case numerically.

\section{Conclusions}

In this paper we investigated how a post-inflationary early dark
energy era can be realized by $F(R)$ gravity, and how to describe
the inflationary era, the early and late-time dark energy era with
a single $F(R)$ gravity model. For our study we used two
approaches, namely a quantitative approach in which we used a
well-known reconstruction technique in order to realize a
generalized Hubble rate that describes all the above mentioned
cosmological eras. In our second approach, we used several
qualitative arguments in order to construct an appropriate $F(R)$
gravity that may realize inflation with the early and late-time
dark energy eras. In the process of selecting a suitable early
dark energy term, we investigated the constraints that the free
parameters of the model must satisfy in order for an early dark
energy era to be realized. Moreover, for one of the proposed
models, we investigated the cosmological behavior of the model at
late times, with and without the early dark energy term. We
numerically integrated the Friedmann equation, by using the
redshift as variable, and we expressed all the physical quantities
as functions of the statefinder function $y_\mathrm{H}(z)$, which
solely depends on the redshift and the Hubble rate $H(z)$. The
results of our numerical analysis indicated that the model is very
similar to the $\Lambda$CDM model at late-times, and also we
demonstrated which values of the free parameters make the early
dark energy term to have no effect on the late-time dark energy
era. Also we investigated the behavior of the growth factor for
the same model, and of the effective gravitational constant
focusing on redshifts $z\leq 10$. For all the cases we studied, we
found that the early dark energy term does not affect at all
neither the effective gravitational constant, nor the growth
factor. With regard to the effective gravitational constant, we
found that for some values of the wavenumber, the gravitational
constant is equal to the present time value of Newton's
gravitational constant, however, there are differences for a
limited set of values of the wavenumber, mainly large values. This
study should be extended to even larger redshifts, and this in
principle could have an observable effect on the CMB, via the
gravitational interaction and the Thomson scattering when the CMB
was generated, exactly on the last scattering surface. In general
the pivot scale dependence of cosmological parameters is a quite
confusing issue in modern theoretical cosmology, so caution is
required on this topic, therefore we did not go deeper in our
analysis. In addition, with regard to the growth factor, since the
latter characterizes the growth of the matter perturbations, an
important quantity is the so-called growth index $\gamma$, which
relates the growth factor $f_\mathrm{g}(z)$ with the matter
density parameter $\Omega_\mathrm{m}(z)$ as
$f_\mathrm{g}(z)=\Omega_\mathrm{m}(z)^{\gamma (z)}$. The form of
the growth index is currently unknown, and its value cannot be
determined directly, however, its value may be inferred by the
observational values of $\Omega_\mathrm{m}(z)$ and
$f_\mathrm{g}(z)$. A cosmographic approach \cite{Benetti:2019gmo}
may also shed some light on the exact functional form of the
growth index, but we defer this study to a future work.

\section*{Acknowledgments}

This work is supported by MINECO (Spain), FIS2016-76363-P, and by
project 2017 SGR247 (AGAUR, Catalonia) (S.D.O). This work is also
supported by MEXT KAKENHI Grant-in-Aid for Scientific Research on
Innovative Areas ``Cosmic Acceleration'' No. 15H05890 (S.N.) and
the JSPS Grant-in-Aid for Scientific Research (C) No. 18K03615
(S.N.).

\end{document}